\documentstyle[aps,epsfig]{revtex}

\begin{document}

\title{Decay of superfluid turbulence via Kelvin wave radiation and vortex 
reconnections.}

\author{M. Leadbeater$^1$, D. C. Samuels$^2$, C. F. Barenghi$^2$, and C. S. Adams$^1$}

\address{$^1$Department of Physics, University of Durham , Durham DH1 3LE.\\
$^2$Department of Mathematics, University of Newcastle, Newcastle NE1 7RU.}

\date{\today}
\draft
\twocolumn
\maketitle

\begin{abstract}

The elementary processes involved in the decay of superfluid turbulence 
in the limit of low temperature are studied by numerical simulations of vortex ring
collisions. We find that small vortex rings produced by reconnections
eventually annihilate in a collision where
all their energy is converted into Kelvin waves and sound. 
We show that sound emission due to Kelvin waves also 
leads to a loss of vortex line length but this dissipation
mechanism alone is too 
small to account for the experimentally observed decay.
 
\end{abstract}


One of the most challenging problems in fluid mechanics is the development of a 
complete hydrodynamical description of turbulence. Progress is often made by
studying fluids where the theoretical description can be simplified. 
For example, low temperature 
superfluids such as liquid helium \cite{qvd} or the recently discovered dilute Bose-Einstein condensates \cite{fermi} are particularly attractive due to the 
dramatic reduction in viscosity. In a superfluid, the 
turbulent state consists of a network of interacting vortices known
as a `vortex tangle' \cite{qvd}. In superfluid 
helium-4 the vortex tangle is strongly coupled to
an interpenetrating normal fluid and the system shares many of the features
of classical turbulence \cite{vine00}. However, in the limit of low temperature,
the normal fluid is negligible, but experiments
still indicate a temperature independent decay of the turbulence
state \cite{davi00}. 
In this case, the emission of sound either by vortex reconnections or 
vortex motion is the only active dissipation mechanism. Similarly, 
the time scale for the formation of vortex arrays in dilute Bose-Einstein condensates
is also temperature independent suggesting that
vortex-sound interactions may also be important in this relaxation
process \cite{abos01}.
In an earlier work we showed that significant sound energy is released
during vortex reconnections \cite{lead01}.
However, at the typical vortex line densities
of superfluid helium-4 experiments \cite{davi00},
or in the regular structures found in a vortex lattice \cite{abos01}
one might expect that vortex reconnections are relatively infrequent,
and that the continuous emission of sound by accelerating vortices may 
be the dominant dissipation mechanism. Unfortunately there is
a scarcity of quantitative predictions of the significance of vortex motion as
a dissipation mechanism in superfluid turbulence. Analytical results for
the sound radiated by moving vortices exist only for simple cases such as a 
co-rotating pair \cite{pism99}. Furthermore, conventional numerical simulations of superfluid turbulence based on vortex filaments governed by incompressible
Euler dynamics (the Biot-Savart law) \cite{tsub00} are unable to describe
sound emission.

An elegant model of quantum fluid mechanics is provided
by the Gross-Pitaevskii (GP) equation. The GP model
represents an extension of Euler fluid
mechanics to include the quantisation of circulation, vortex 
core structure and vortex-sound interactions. Although the GP model
does not accurately represent the physics of superfluid
helium-4, it has been shown to provide qualitative and sometimes quantitative insight
into the critical velocity for vortex nucleation \cite{fris92,wini00},
vortex line reconnections \cite{kopl93}, vortex ring collisions
\cite{kopl96}, the decay of superfluid turbulence \cite{nore97},
and sound emission due to vortex reconnections \cite{lead01}.
 
To study sound emission due to vortex motion we excite 
Kelvin waves on a vortex ring and measure the length of the ring
as a function of time. 
The Kelvin waves are produced by colliding two or more vortex rings.
The initial state is constructed from
stationary vortex ring solutions of the uniform flow equation 
found by Newton's method \cite{wini99b}. The
desired configuration is obtained by multiplying
the individual vortex ring states. This initial 
state is then evolved according to the dimensionless GP equation,
\begin{equation}
i\partial_t\psi=-\textstyle{1\over 2}\nabla^2\psi+(\vert\psi\vert^2-1)\psi~,
\end{equation}
using a semi-implicit Crank-Nicholson algorithm. 
In dimensionless units, distance and velocity are 
measured in terms of the healing
length, $\xi$, and the sound speed, $c$, respectively. 
In addition, the asymptotic number density, $n_0$, is rescaled to unity.
The computation box with volume, $V=(50)^3$, is divided into 
$10^6$ grid points with a spacing of $0.5$. A grid spacing of 0.25 was also used to
test the accuracy of the numerical methods. The time step is $0.02$
and a typical simulation is run for $1.5\times10^5$ steps.
Simulations have been performed with two, three and
four vortex rings with radii ranging from 2.86 to 18.1 healing lengths.
To convert the dimensionless 
units into values applicable to superfluid helium-4, we take the number density as $n_0=2.18\times10^{28}$~m$^{-3}$, the quantum of circulation as 
$\kappa=h/m=9.92\times10^{-8}~{\rm m}^2{\rm s}^{-1}$,
and the healing length as $\xi/\sqrt{2}=0.128$~nm \cite{ray64}.
This gives a time unit $2\pi\xi^2/\kappa=2$~ps, and therefore 
for superfluid helium-4 our simulations would correspond to a real time of 6~ns.
Whereas for a sodium vapour condensate with $\xi\sim 0.2~\mu$m \cite{abos01}, the time
unit is $\sim 15$~ns giving a simulation time of approximately $45~\mu$s. 

In the first example we consider a collision between a large and a small
vortex ring. Such collisions are important in vortex tangle dynamics
because small vortex rings are often produced in collisions between
larger structures ~\cite{kopl96}. A sequence of density isosurface plots illustrating 
the collision are shown in Fig. 1
\footnote{The original animated movies for figures 1 and 4 can be found
 at http://massey.dur.ac.uk/ml/.}.
Both vortex rings propagate in the positive $x$ direction with
their propagation axes offset by the radius of the larger ring, $R_1=18.1$. The collision occurs at $t=110$,
and leads to the destruction of the small vortex ring. The 
vortex energy is converted into a sound pulse and Kelvin waves.
The sound pulse propagates away from the large vortex ring and
appears as a density minimum at $t=120$ in Fig.~1.
The Kelvin waves appear as two kinks which propagate around the  
vortex ring in opposite directions.
 
\begin{figure}[hbt]
\centering
\epsfig{file=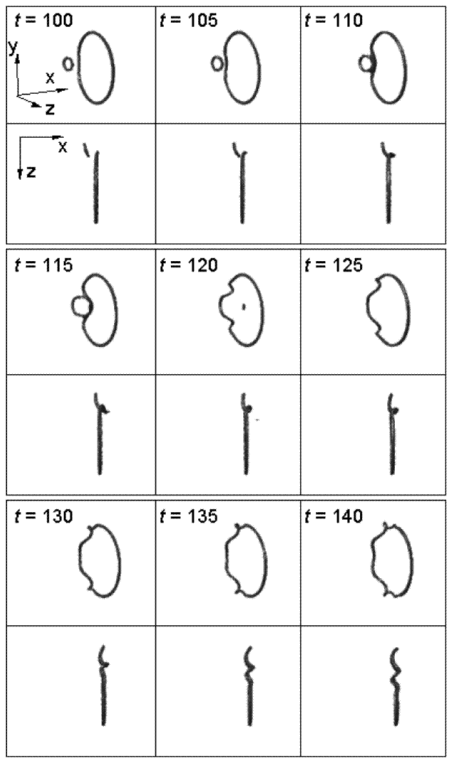,width=7.5cm,angle=0}
\caption{Sequence of density isosurfaces ($\vert\psi\vert^2=0.35$) 
illustrating a vortex ring collision with ring radii $R_1=18.1$ and $R_2=2.86$.
A top view is shown below each frame. The rings collide at $t=110$. The
reconnection produces a sound pulse which appears as a small dot in the $t=120$ frame.
}
\label{fig:1}
\end{figure}

In Fig. 2 we plot the vortex line length, $\ell$, as a function of time.
The vortex line length is evaluated numerically by searching for phase
singularities, estimating the length of vortex line within a grid
cube and summing over all cubes. We have checked the accuracy of the method by
performing the same calculation with a grid spacing of 0.25 and 0.5.
The two calculations agree to within a fraction of the healing length. 
During the reconnection at $t=110$, the vortex line 
length is stretched from its initial value of $2\pi(R_1+R_2)\sim 130$,
and then decreases dramatically due to the conversion of energy into
sound \cite{lead01}. After the collision the small vortex
ring disappears completely while the mean
radius of the large vortex ring is slightly larger, $R_1'=18.8$. 

The vortex line length oscillates due to the motion of the Kelvin waves around the 
vortex ring. 
By fitting the oscillations to a sine wave (shown bold in Fig.~2) we find 
that the period is $134\pm 1$. This corresponds to
a Kelvin wave velocity of 0.44 times the speed of sound. The 
dispersion relationship for Kelvin waves is \cite{donn91}
\begin{equation}
\omega=\frac{\kappa}{4\pi}k^2\ln\left(\frac{2}{ka}\right)~,
\label{eq:disp}
\end{equation}
where in dimensionless units, $\kappa=2\pi$ and the 
core size $a=1/\sqrt{2}$, therefore the 
group velocity of a Kelvin wave with center wavelength $\lambda$
is
\begin{equation}
v_{\rm g}=\frac{\pi}{\lambda}\left[2\ln\left(\frac{\sqrt{2}\lambda}{\pi}\right)
-1\right]~.
\label{eq:v_g}
\end{equation}
A velocity of 0.44 corresponding to a Kelvin wavelength of 5.3 healing lengths,
which agrees with a measurement made from Fig.~1. In this intermediate 
wavelength region, the Kelvin wave group velocity (\ref{eq:v_g}) is only 
weakly dependent on wavelength ($ 0.4\leq v_{\rm g}\leq 0.63$ for $5\leq\lambda\leq 35$). 
Consequently, the dispersion of the 
Kelvin wave packet is relatively slow.

During the Kelvin wave oscillations the mean vortex line
length gradually decreases. It is interesting
to compare the observed length decrease with that expected for a 
Kelvin wave cascade, where energy is transfered to 
shorter and shorter wavelengths with a cut-off below
a critical wavelength. In this case the vortex line
density $L=\ell/V$, can be described by an equation of the form \cite{vine57},
\begin{equation}
\frac{{\rm d}L}{{\rm d}t}=-\frac{\kappa}{2\pi}\chi_2 L^2~,
\label{eq:vine}
\end{equation}
where $\kappa=2\pi$ in dimensionless units, and $\chi_2$ is
a dimensionless coefficient. A linear fit to $1/L$ using the 
data from Fig. 2 for times between $t=500$ and $t=2900$ gives a
coefficient, $\chi_2 \sim 0.006\pm0.001$. 
Both superfluid turbulence experiments \cite{vine57} and
numerical vortex dynamics simulations where an 
energy cut-off is assumed \cite{tsub00} suggest a 
coefficient, $\chi_2\sim 0.3$, in the limit of low temperature. 
This suggests that Kelvin wave radiation alone is not sufficient
to account for the decay of superfluid turbulence.

\begin{figure}[hbt]
\centering
\epsfig{file=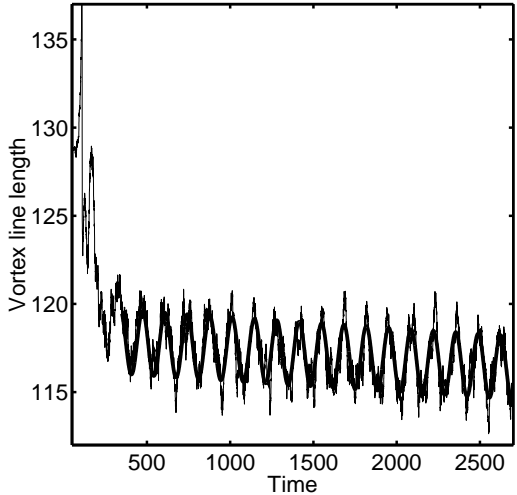,width=7cm,clip=,bbllx=80,bblly=175,bburx=500,bbury=590}
\caption{The vortex line length, $\ell$, as a function of time for 
the collision shown in Fig.~1. During the 
reconnection there is a sudden loss of line
length due to the emission of a sound pulse. 
Subsequently the length decays due to Kelvin wave radiation.
The bold line is a fit to a function of the form 
$\ell=\ell_0[1+a\sin(2\pi t/T+\phi)]/(1+\chi_2 \ell_0 t/V)$, where
the period of the oscillations, $T=134\pm 1$, corresponds to the time
for the Kelvin wave to go half way around the vortex ring.
A linear fit to $1/L$ against $t$ between $t=500$ and $t=2900$ determines the
coefficient in Eq.~(\ref{eq:vine}), $\chi_2 \sim 0.006\pm0.001$. 
This data is for a grid spacing of 0.25 healing lengths.}
\label{fig:2}
\end{figure}

To test the generality of this result we repeated the calculation 
for different initial conditions. First, a collision
involving initial vortex rings with radii $R_1=18.1$ and $R_2=5.74$
produces the decay shown in Fig.~3. 
In this case the smaller vortex ring is not completely destroyed in
the initial collision. Instead after making
repeated reconnections with the main vortex ring between $t=400$ and $t=1250$,
it annihilates at $t\sim 1250$ leaving only the larger ring.
The large vortex ring then decays by Kelvin wave radiation 
at a rate consistent with the results in Fig.~2.  
Second, collisions involving three vortex rings 
reproduce the same characteristic decay with large drops in 
the vortex line length during reconnection events interspersed
with gently sloping regions of Kelvin wave radiation.
All the simulations show that the Kelvin wave decay coefficient is 
independent of the initial condition.
This can be explained by the fact that the radiation spectrum 
is dominated by short wavelength Kelvin waves \cite{kelvin}.

The length of vortex line destroyed during a reconnection is
strongly dependent on the reconnection angle \cite{lead01}.
Consequently, to model a completely disordered vortex tangle we need
to simulate a system which samples a complete
range of reconnection angles. The simplest case where
this is realised is four initial vortex rings,
propagating inwards, as in the first frame of Fig. 4. 
The propagation axes are offset by one healing length to break the symmetry
(the symmetric case completely annihilates in the first few hundred time units). 
A similar system, although with a very different length scale, 
has been studied using a classical vortex filament calculation \cite{vass01}. 
The decay of the vortex line length is shown in Fig. 5. 
The initial vortex line density is much 
larger than in the previous examples, therefore there are a larger number of 
reconnections with a broad distribution of reconnection 
angles. A sequence of reconnections before $t\sim 800$ results in the formation of two
large vortex rings which avoid each other until $t\sim2600$. 
Again the decay coefficient for Kelvin wave radiation is consistent with 
Fig.~2.  The collision at $t\sim 2600$ sets up another phase of repeated reconnections 
producing smaller vortex rings which annihilate in collisions with
larger structures. This scenario of vortex tangle decay involving the production of 
small vortex rings is also observed in the classical
vortex filament calculations of Tsubota {\it et al}. \cite{tsub00}.
Our main conclusion that Kelvin wave radiation alone cannot account 
for the observed decay of superfluid turbulence is illustrated by plotting 
a decay characterised by $\chi_2=0.3$ in Fig. 5.  

\begin{figure}[hbt]
\centering
\epsfig{file=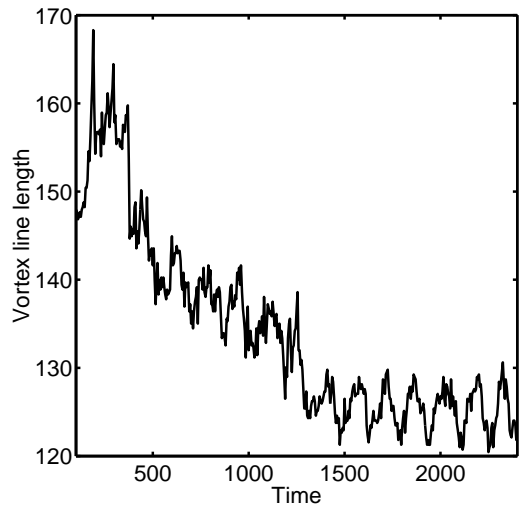,width=7cm,clip=,bbllx=80,bblly=175,bburx=500,bbury=590}
\caption{The vortex line length as a function of time for 
a collision between vortex rings with radii $R_1=18.1$ and $R_2=5.74$. 
The initial increase in the vortex line length 
is due to stretching during the first collision. Subsequently there are
a large number of reconnections which result in the production of
smaller rings. After $t=1100$ a sequence of reconnections
leaves only one vortex ring which then decays by Kelvin wave radiation. 
The decay between $t=1500$ and $t=2500$ is consistent with 
the coefficient found previously, $\chi_2\sim0.006$.}
\label{fig:3}
\end{figure}

In summary, we have studied the basic processes involved in 
the decay of superfluid turbulence by numerical simulation of vortex ring
collisions. At high vortex line densities our simulations 
suggest that the primary mechanism for the loss of vortex line length
is sound emission during reconnection events. 
The magnitude of the loss due to reconnections is in stark contrast 
to the relatively slow decay due to sound
radiation from Kelvin waves. We conclude that Kelvin wave
radiation alone is not sufficient to account for
the decay of superfluid turbulence but that Kelvin waves plus vortex reconnections
could be.

\begin{figure}[hbt]
\centering
\epsfig{file=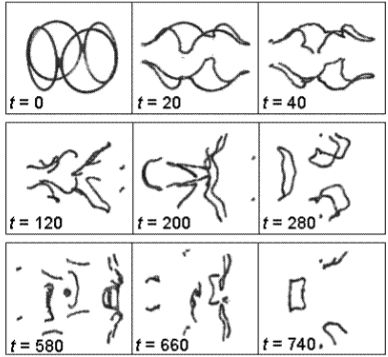,width=7cm,angle=0,clip=,bbllx=70,bblly=305,bburx=485,bbury=700}
\caption{Sequence of density isosurfaces ($\vert\psi\vert^2=0.35$) 
illustrating a collision involving
four converging vortex rings each with radius $R=18.1$.
A large number of reconnections result in the generation of small 
structures which annihilate into sound energy producing only two 
vortex rings by $t=740$. Note that because of the periodic boundary conditions
the loop structures connect on opposite faces of the numerical box. }
\label{fig:4}
\end{figure}

\begin{figure}[hbt]
\centering
\epsfig{file=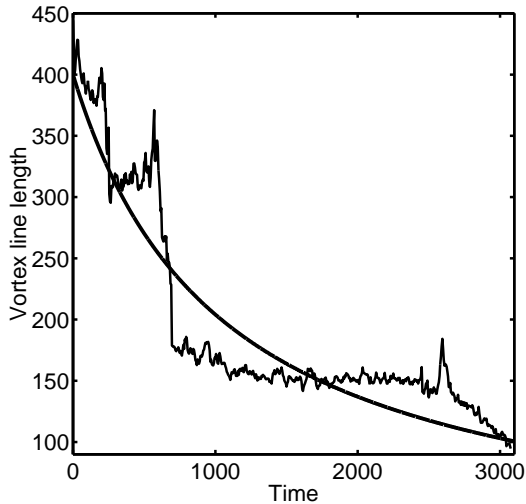,width=7cm,angle=0,clip=,bbllx=80,bblly=175,bburx=500,bbury=590}
\caption{The decay of vortex line length (thin line) for the collision involving
four converging vortex rings shown in Fig. 4.
For $t<800$ there are a large number of reconnections resulting
in a dramatic decrease in length. After $t\sim 800$ only two rings
remain and there are no more reconnections until $t=2400$. In this
region the loss of vortex line length follows the prediction for
Kelvin waves with a decay coefficient consistent with the data
from Figs. 2 and 3. A decay characterised by $\chi_2=0.3$ is plotted
(thick line) to illustrate the difference between this case and the much slower 
decay rate due to Kelvin wave radiation only.}
\label{fig:5}
\end{figure}

\acknowledgements
Financial support was provided by the EPSRC.


\begin{references}
\bibitem{kelvin} The power radiated is proportional to a 
high power of the Kelvin wave frequency \cite{vine00}. 
From equation (\ref{eq:disp}) one finds that the power spectrum is 
sharply peaked around a wavelength of order 5 healing lengths. 
As the spectrum of Kelvin waves produced in a typical vortex ring collision
is also peaked around 5 healing lengths, see Fig. 1, the decay is dominated
by this wavelength region.
\end{references}
\end{document}